%% file: main.tex
\documentclass[authorversion,manuscript,screen,sigconf]{acmart}

\setcopyright{acmcopyright}
\copyrightyear{2021}
\acmYear{2021}
\acmDOI{10.31219/osf.io/nu9p3}

\acmConference[EHPHCI 2021]{ACM CHI Workshop on Esports and High Performance Human Computer Interaction 2021}{May 08--13, 2021}{Yokohama, Japan}
\acmBooktitle{EHPHCI 2021: ACM CHI Workshop on Esports and High Performance Human Computer Interaction,
May 08--13, 2021, Yokohama, Japan}
\usepackage[ruled, vlined]{algorithm2e}
\usepackage{wrapfig}
\usepackage{todonotes}
\begin{document}

\input{title-abstract.tex}



% \keywords{datasets, neural networks, gaze detection, text tagging}

%\input{teaser.tex}

\maketitle

\input{body.tex}

\bibliographystyle{ACM-Reference-Format}
\bibliography{main}

\end{document}

%% file: title-abstract.tex
% \title{First Person Aiming at Low Latency for Esports}
\title{A Case Study of First Person Aiming at Low Latency for Esports}

\author{Josef Spjut}
\email{jspjut@nvidia.com}
\orcid{0000-0001-5483-7867}
\affiliation{%
  \institution{NVIDIA}
  \country{USA}
}
\author{Ben Boudaoud}
\email{bboudaoud@nvidia.com}
\affiliation{%
  \institution{NVIDIA}
  \country{USA}
}
% \author{Seth Schneider}
% \email{sschneider@nvidia.com}
% \affiliation{%
%   \institution{NVIDIA}
\author{Joohwan Kim}
\email{sckim@nvidia.com}
\affiliation{%
  \institution{NVIDIA}
  \country{USA}
}
\renewcommand{\shortauthors}{Spjut et al.}

\renewcommand{\shorttitle}{A Case Study of First Person Aiming at Low Latency for Esports}

\begin{abstract}
% When a computer system has lower input-to-output latency, users should be able to complete (some) tasks more quickly.
Lower computer system input-to-output latency substantially reduces many task completion times. 
%In fact, the literature shows that improvements in Fitts' Law and aiming task performance often exceed the pure latency reduction.
In fact, literature shows that reduction in targeting task completion time from decreased latency often exceeds the decrease in latency alone.
However, for aiming in first person shooter (FPS) games, some prior work has demonstrated diminishing returns below 40 ms of local input-to-output computer system latency.
In this paper, we review this prior art and provide an additional case study with data demonstrating the importance of local system latency improvement, even at latency values below 20 ms.
Though other factors may determine victory in a particular esports challenge, ensuring balanced local computer latency among competitors is essential to fair competition.
\end{abstract}

%% file: body.tex
\section{Introduction}
In the world of esports every possible advantage is sought out in order to beat the competition. 
Competitive gamers play games to win, and get most of their enjoyment from doing well. 
First Person Shooter (FPS) games like Counter-Strike, Overwatch and Valorant allow players to compete at many skill levels in online ladders to see who will come out on top. 
Market trends and online message boards suggest that competitive gamers seek the highest refresh rates and lowest latency, constantly pursuing every minor advantage they might provide. 
In this study, we investigate the impact of local system latency on FPS aiming performance, with a particular focus on the low end of latency that most interests competitive FPS gamers.
\begin{figure}
    \centering
    \includegraphics[width=\columnwidth]{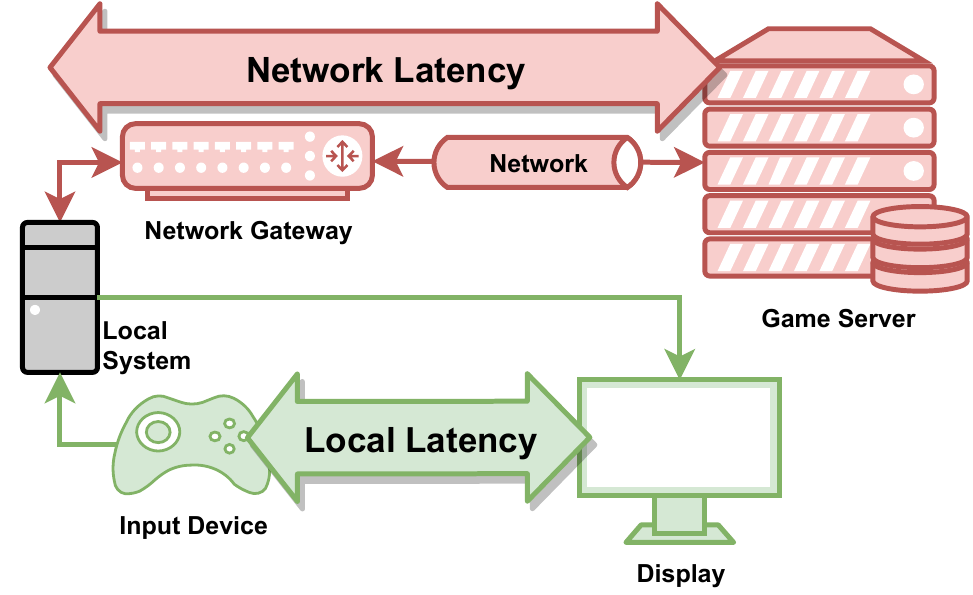}
    % \vspace{-3mm}
    \caption{Diagram outlining the difference between \emph{local} latency, the focus of this work, and \emph{network} latency, the more commonly referred to term historically. Note that many in-game events are handled at the local latency, without game server intervention.}
    % \vspace{-4mm}
    \label{fig:LocalLatency}
\end{figure}

\subsection{Local System Latency}

In computing systems, there are many types of latency that contribute to the total time it takes for a user's actions to produce output from the system. 
A commonly discussed latency in esports and online games is the network latency.
Network latency is important because it describes how long it takes for local actions to be received at the server, and subsequently, how long it then takes to deliver those actions to each of the connected players.
However, in this work, we focus on the \emph{(local) system latency}, that is the time from a user's input until the result of that input is delivered by a computer.
In this study, we measured system latency as the time for a mouse click to cause a pixel change in the display, and report the average of many measurements in each condition.
Note that while we report the average for our latency values, the actual measured latencies occur over a range surrounding that average, and furthermore, other latencies (such as mouse sensor input, or audio output) were not measured and are assumed to be different by some relatively fixed offset from the measured values.

\section{Background}

\begin{figure}
    \centering
    \includegraphics[width=\columnwidth]{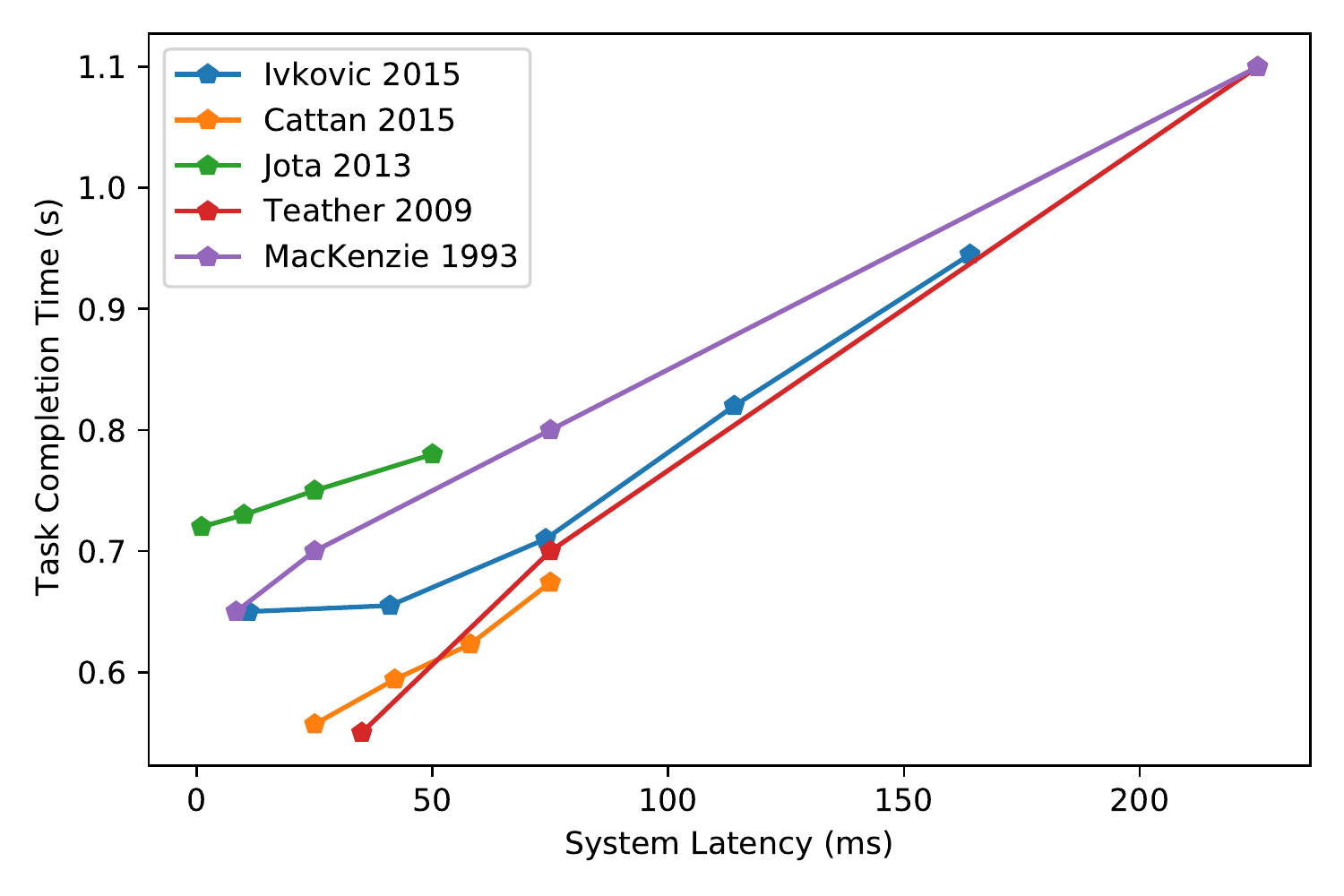}
    % \vspace{-6mm}
    \caption{The effect of local system latency on completion time for simple pointing tasks from previous publications. Trial data was selected for index of difficulty between 2 and 2.5 bits except for the Jota data, which has an index of difficulty of 1.58 bits.}
    \label{fig:related_aiming}
    % \vspace{-4mm}
\end{figure}

A number of prior studies consider pointing and aiming tasks in the context of a computer system’s added latency.
It has been observed that, when the latency differences inherent to frame rate changes are controlled for, latency reduction offers a more significant aiming task performance benefit than frame rate \cite{spjut2019latency}.
We collected data from Ivkovic~\cite{ivkovic2015quantifying}, Cattan~\cite{cattan2015reducing}, Jota~\cite{jota2013howfast}, Teather~\cite{teather2009effects}, and MacKenzie~\cite{mackenzie1993lag} in an effort to visualize how latency may change aiming task completion time, similar to how target size and distance are known to affect task performance via a combined metric known as "index of difficulty" (ID)~\cite{fitts1964information}. 
% Ben's attempt to summarize prior art referenced above
These surveys are interesting in that they cover a diverse set of interfacing modalities.
Jota \cite{jota2013howfast} and Cattan \cite{cattan2015reducing} look at touch-based, dragging interfaces over different ranges of latency.
Teather \cite{teather2009effects} and MacKenzie \cite{mackenzie1993lag} both look at pointer-based mouse interfaces, but MacKenzie considers traditional 2D UI pointing tasks, while Teather focuses on completing 3D visualized spatial tasks.
Ivkovic \cite{ivkovic2015quantifying} also looks at mouse-based interaction, but for non-pointer FPS aiming tasks, similar to the focus of our study.
We include FPS aiming alongside other aiming tasks since Looser~\cite{looser2005validity} shows that FPS aiming follows the same Fitts' law relationship as other pointing tasks. 
In Figure~\ref{fig:related_aiming}, you can see results selected from these studies with a similar ID to make it more reasonable to compare them on the same plot.

The apparent trend in this related work is that task completion time increases (gets worse) as latency increases. 
In all studies shown, this trend shows continued improvements all the way toward 0 latency. 
The Ivkovic \cite{ivkovic2015quantifying} study is interesting, as it is the only one using a FPS aiming task similar to our work, and shows diminishing returns from latency reduction below about 50 ms. 
However, this work's lowest-latency data point represents a change in the V-SYNC setting, which may be a confounding factor. 
More study around these low latency levels would certainly augment this limited understanding.

\section{Comparing 12 and 20 ms}

\begin{figure}
    \centering
    \includegraphics[width=\columnwidth]{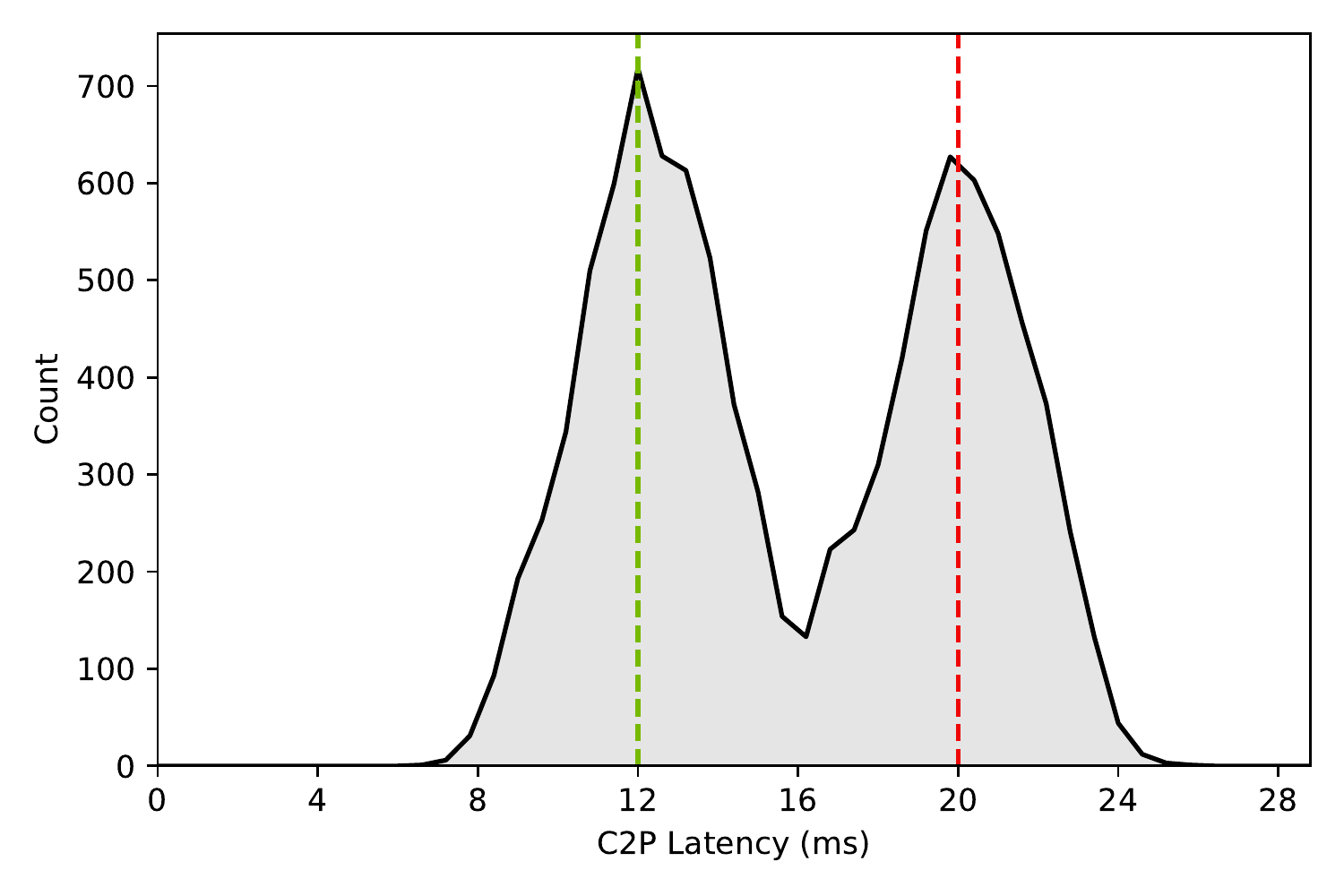}
    % \vspace{-6mm}
    \caption{Click to photon distribution from a selection of clicks in our study across both 12 ms and 20 ms conditions. As expected, the two peaks are mostly separated, though there is some overlap in the middle.}
    \label{fig:c2p_distribution}
\end{figure}

Published research, including that found in Figure~\ref{fig:related_aiming}, has often struggled to study latency below 20 ms.
For this reason, we endeavor to study exclusively 20 ms and below in this work. 
In developing this study, we found that the lowest system latency we were able to achieve consistently on our test systems was 12 ms.
Through further testing injecting additional latency, we found that by adding 8 ms of latency on average, the measured latency distributions only slight overlap (see Fig.~\ref{fig:c2p_distribution}).
Thus we selected 12 ms as the lowest latency condition for our experiment, and 20 ms as a condition sufficiently low, yet high enough to show distinct results from the 12 ms condition.

In order to get the minimum average latency to 12 ms, we selected appropriate hardware and software tools. 
G-SYNC monitors with 240 Hz refresh rate and fast pixel response time were a key component of this low latency system.
We also use Logitech G203 wired mice, which were measured to consistently have relatively low click-to-USB packet latency.
The computer hardware included Intel Core i7-9700K CPUs and NVIDIA RTX 2080 Ti GPUs.
Finally, we selected the First Person Science (FPSci)~\cite{Spjut19FPSci} platform we designed previously, giving us full control of the inner loop of the application and allowing us to optimize this loop for minimal input-to-visual latency.
FPSci was created to support customized first person aiming experiments such as the one we designed for this work.
This work uses FPSci release v20.07.01\footnote{Experiment configuration at \url{https://github.com/jspjutNV/latencyExpEHPHCI21}.}.

\begin{figure}[b]
    \centering
    \includegraphics[width=\columnwidth]{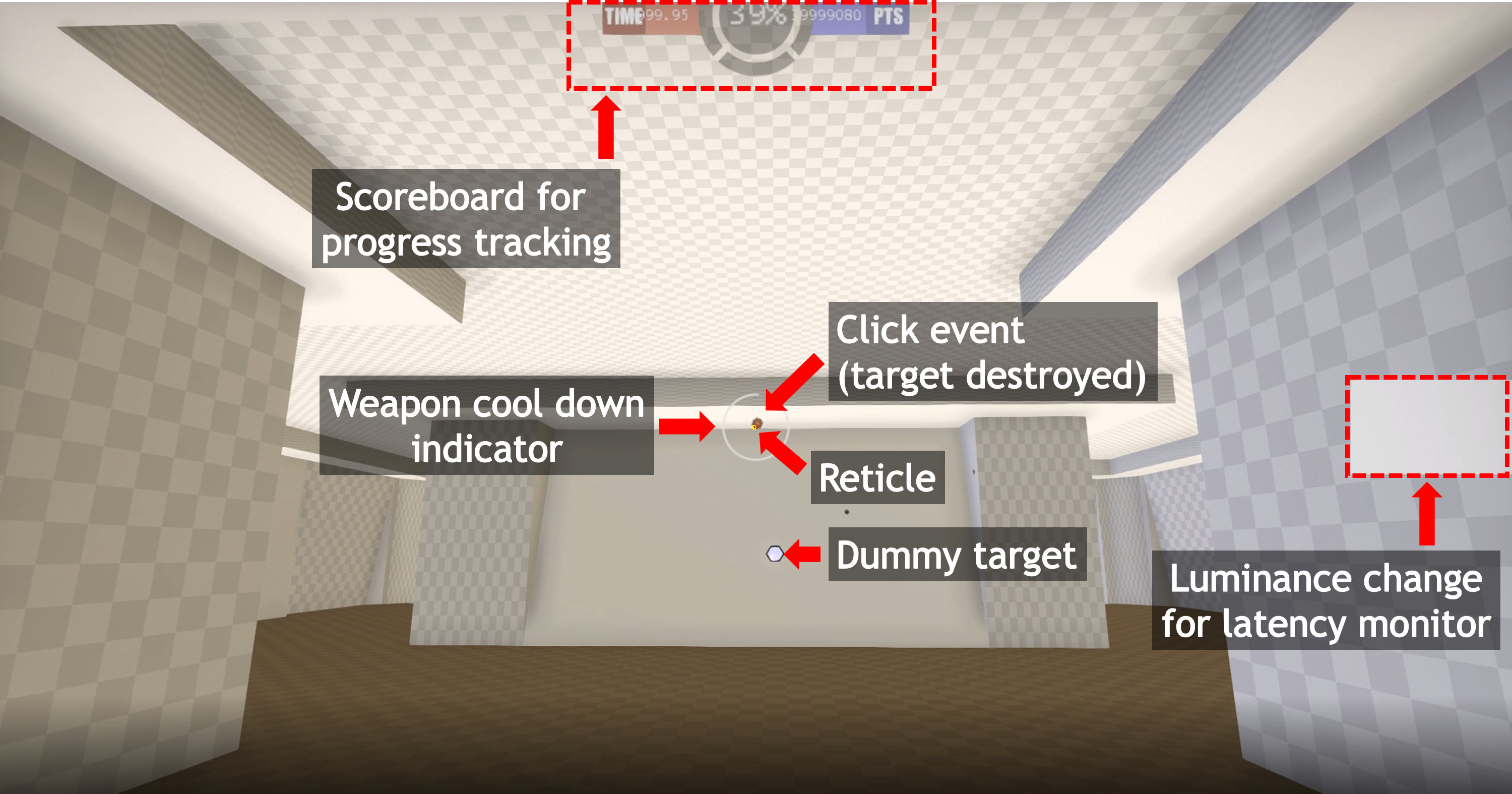}
    \caption{Annotated in-app user view just after the destruction of a target showing the dummy target (in white).}
    \label{fig:FPSciScreenshots}
    % \vspace{-4mm}
\end{figure}

Our experiment consisted of a simple repeated FPS aiming task evaluated based on completion time. 
Subjects begin the task by centering their aim direction indicator, or reticle, on a (white) reference target (as shown in Figure \ref{fig:FPSciScreenshots}) then press shift on the keyboard to destroy this target.
Once the reference target is destroyed, a test target appears at a randomized position within a world space bounding box, moving along a linear path with a constant velocity.
Periodically (once every 0.8-1.5 seconds) this test target changes direction and velocity.
If the test target leaves its world space bounding box its velocity is reflected about the bounds producing a "bounce" behavior and keeping the target in a desired range of distance and visual angle.
Subjects attempt to align their reticle with the target and click, destroying the target.
If the subject clicks without the reticle over the target a miss is registered and the trial continues.
Following a missed shot a weapon cooldown penalty was assessed, not allowing the user to fire for another 0.5 seconds.
This penalty was included to mimic various game weapon cooldown periods for similar styles of weapons, and has the effect of amplifying missed shots in terms of task completion time.
Comparing a 0.5 second cooldown to no cooldown is outside the scope of this work.

We distributed our experiment to 8 users, each of which completed the aiming task 400 times at 12 ms latency and another 400 times at 20 ms latency (with 8 ms of added delay). 
These 800 trials were spread across 8 different sessions after a 20 trial warm up session was completed.
During the warm up session each player could customize and select their preferred mouse sensitivity setting.
Each of the 100 trial sessions, including the warm up, were broken up into blocks of 10 trials. 
The players were given brief breaks of at least 1 second after each block, and allowed a longer break between sessions if they desired.
% 12 ms was the lowest average latency we were able to achieve with our experimental configuration (RTX 2080 Ti) at 240 Hz and is a measurement of the average click to photon time as reported by the latency analyzer. 

\begin{figure}
    \centering
    \includegraphics[width=\columnwidth]{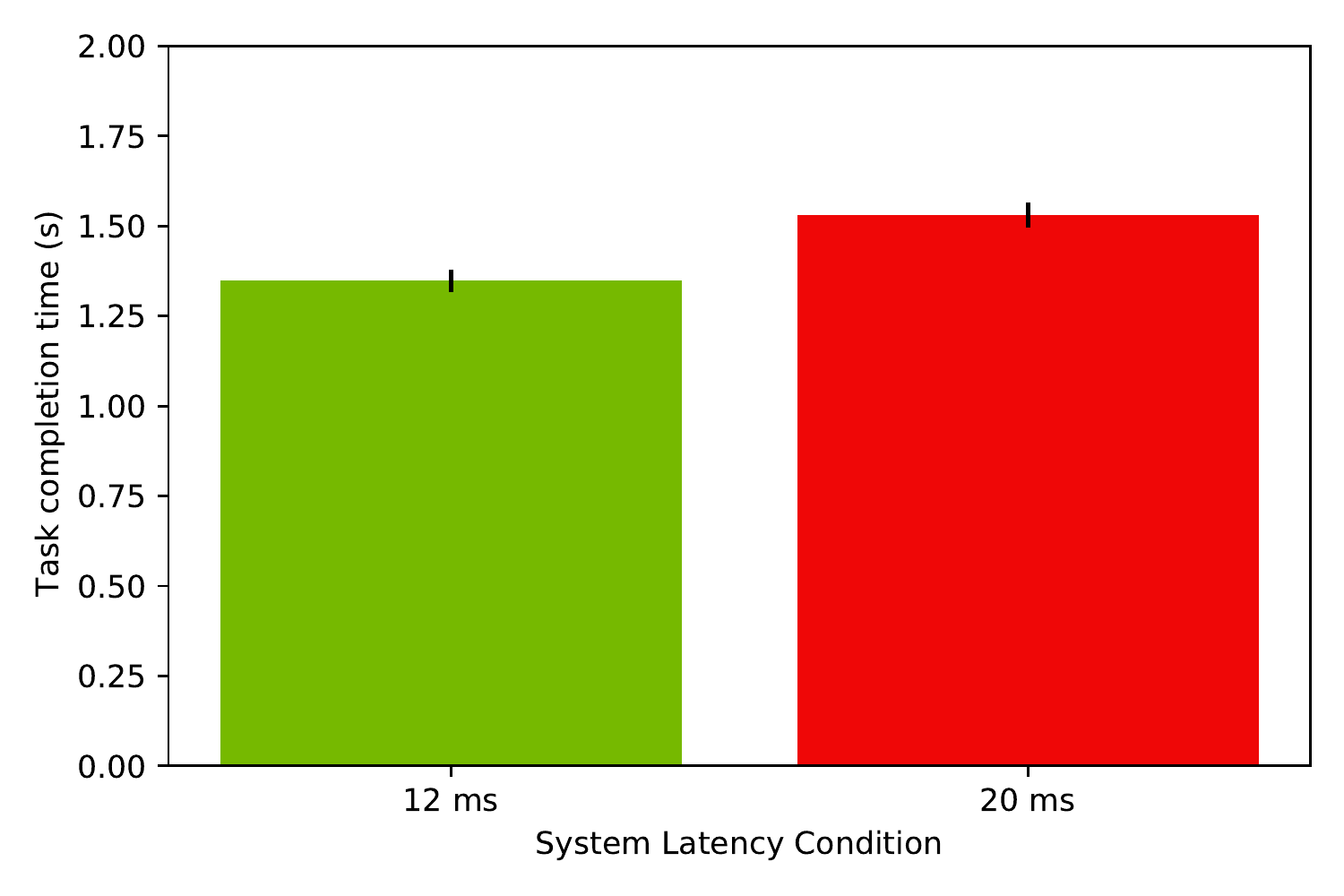}
    % \vspace{-6mm}
    \caption{Median aiming task completion time and standard error metric for 3200 trials at 12 ms and 20 ms of system latency.}
    \label{fig:summary20vs12}
    % \vspace{-6mm}
\end{figure}

\begin{figure}
    \centering
    \includegraphics[trim=10 0 10 20, clip,width=\columnwidth]{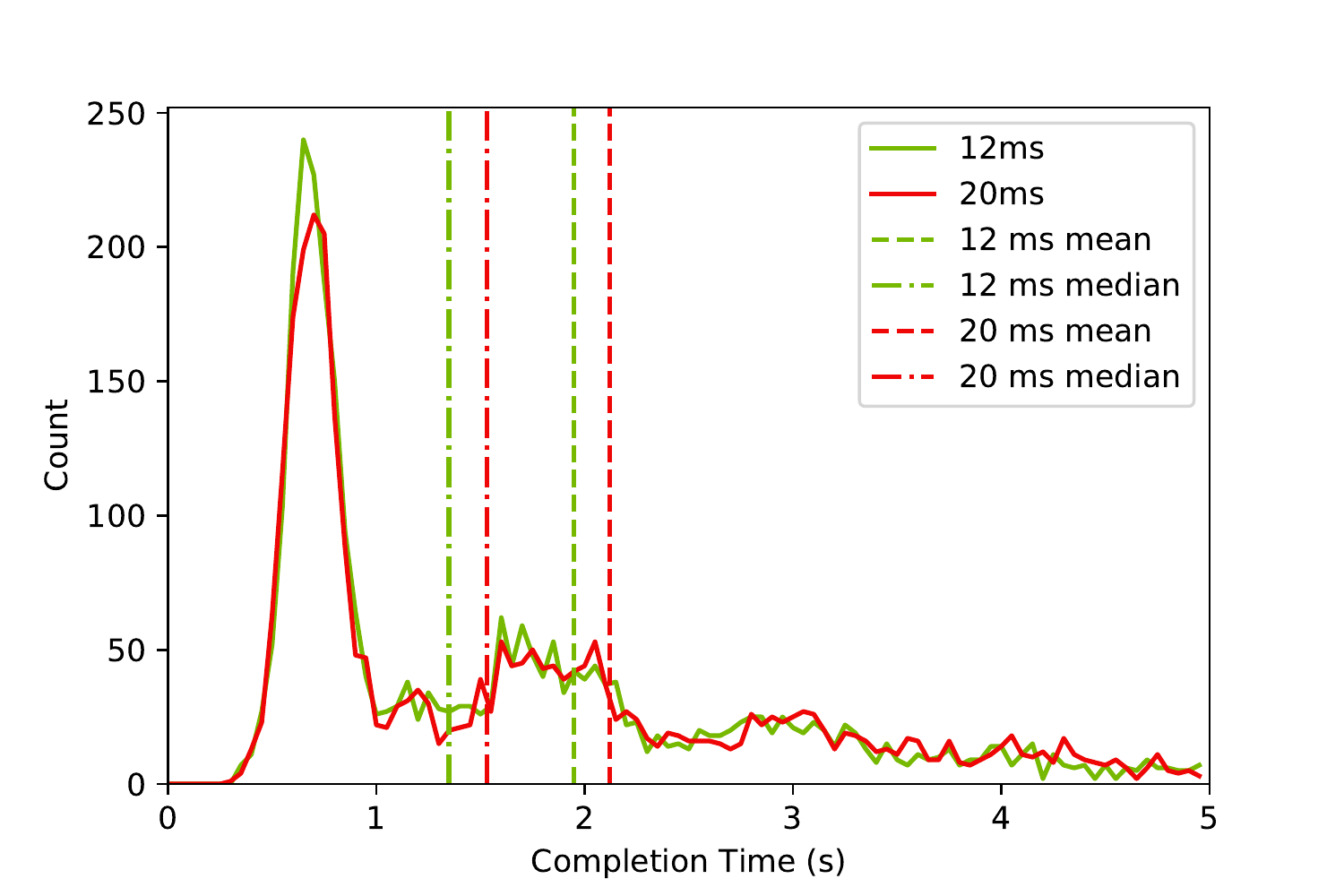}
    % \vspace{-6mm}
    \caption{A histogram showing the completion time distribution for the 12 and 20 ms latency conditions. Means are indicated by the dashed vertical lines and medians by the alternating dotted/dashed lines. Note that the plateau between 1.5-2.2s is likely caused by the 0.5s weapon cooldown and may loosely represent the second shot clustering. }
    \label{fig:CompTimeHistogram}
    % \vspace{-4mm}
\end{figure}

We choose to analyze our results primarily using median task completion time, as the median is often more representative of a \emph{typical} performance of a subject and robust to individual extreme values than mean.
The underlying completion time distributions, with mean and median values plotted is provided in Figure \ref{fig:CompTimeHistogram}.
The median measured completion time for this task across all 3200 trials is lower for 12 ms of latency (1.348 s) than for 20 ms of latency (1.530 s).
These medians are shown with the corresponding standard error metric in Figure \ref{fig:summary20vs12}. 
Consistent with prior art the difference in median task completion time (182 ms) far exceeds the reduction of latency (8 ms).
A Wilcoxon signed rank test (over all subject data) shows that the medians are significantly different (p-value < 0.005). 
We use Cohen's D to examine effect size for the difference in mean completion time between the 12 and 20 ms latency conditions as well.
The overall Cohen's D value for this difference of means is 0.319 indicating a small-to-medium effect.

We present the distribution of task completion time across all subjects by latency condition in Figure \ref{fig:CompTimeHistogram}.
Note that though there is a significant difference in mean and median between the two distributions, the mode values do not differ as substantially.
In addition, the difference between mean and median for the distributions at 12 and 20 ms remains fairly consistent.
This is common for such heavy-tailed task completion time distributions wherein the tail weight (or shape) tends to account for much of the difference in average task completion time.

A keen observer will notice that the task completion times in our experiment are higher than the results for the simple pointing tasks from the prior research publications. 
This is because the task we asked our users to perform was more difficult. 
FPS aiming requires detecting a change or target, determining where the target is, planning for perceived target motion, moving the fingers, hand, arm and wrist to position the aiming reticle over the target, and finally clicking the mouse button. 
This sequence of actions can be completed quite rapidly for those who are experienced and plan well, but when the target has complicated shapes and motion characteristics, players may need to continually update their aim to click on the target.
The process of updating the aim while a target is in motion may be a part of the underlying reason for the increase in task completion time beyond the difference in system latency.

\section{Individual Results}

\begin{figure}
    \centering
    \includegraphics[width=\columnwidth]{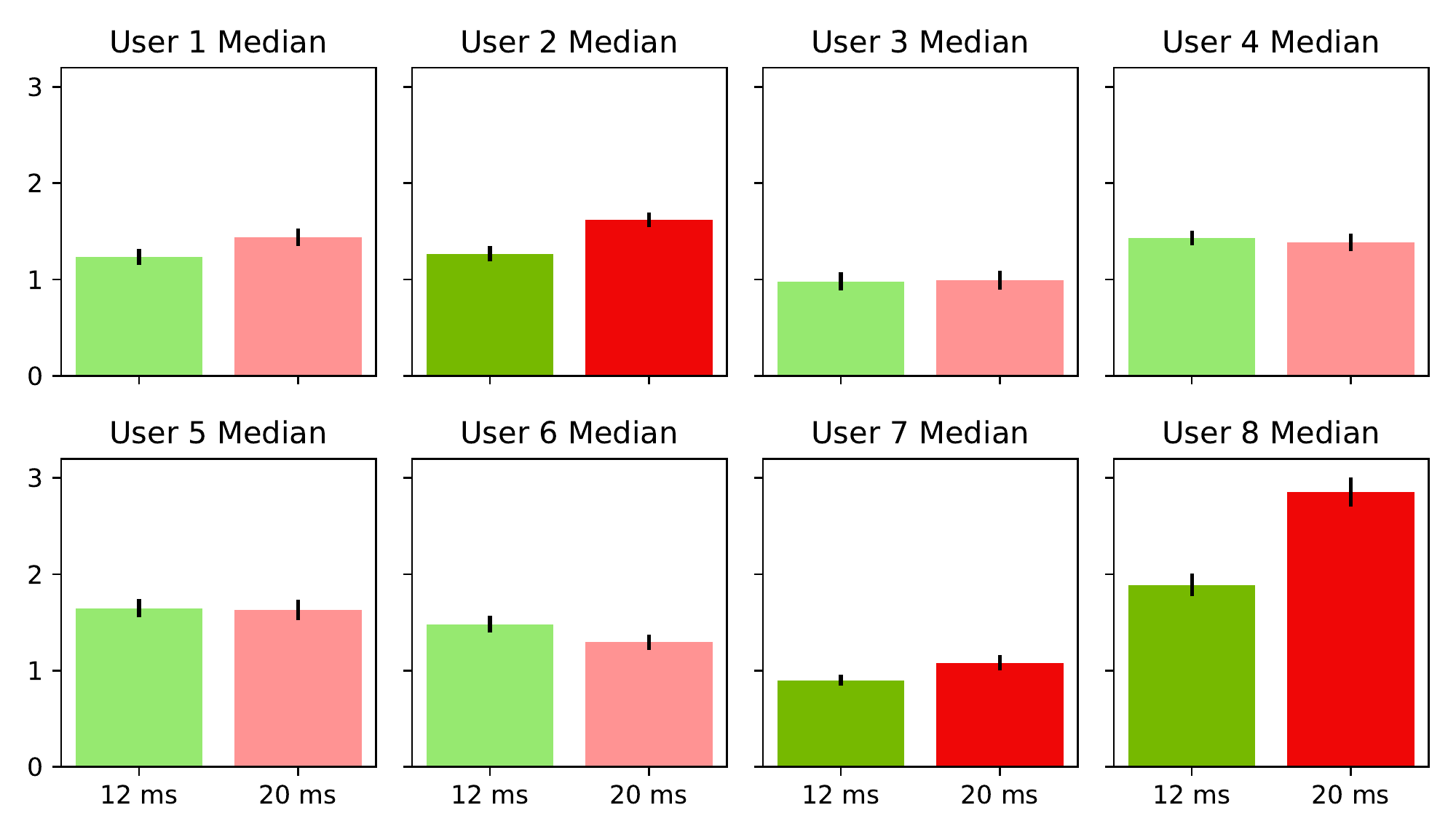}
    % \vspace{-6mm}
    \caption{Per-user median aiming task completion time and standard error metric for 400 trials each at 12 ms and 20 ms of system latency. Users whose results did not reach statistical significance are plotted in the lighter shade.}
    \label{fig:peruser20vs12}
    % \vspace{-4mm}
\end{figure}

While the overall data is interesting on its own, it is also interesting to consider the results of individual users. Figure~\ref{fig:peruser20vs12} plots these individual results. 
Three users saw statistically significant improvements at 12 ms of local latency compared to 20 ms. 
However, the remaining five users saw no statistically significant difference in this experiment. 
This suggests that there may be an individual sensitivity to impacts of latency. 
This sensitivity may be a result of any number of factors including skill level, ambient light conditions, previous practice or experience at low latency, and other  factors. 
Further study is needed to investigate which, if any, of the many possible causes for this individual variation is to blame.

Furthermore, it is worth noting that though we find statistical significance for the claim of benefits from minor latency reduction in this study, high sample count is necessary for reaching significance for this claim. 
An alternative analysis approach wherein per-user medians are compared between latency conditions found no statistically significant difference between the means of these sets of per-user median task completion times.
This is due to the significantly reduced sample size when considering user count (8) to trial count (3200 or 400 per user).

We are reluctant to draw too many strong conclusions about individual variation given the small pool of users, and the lack of additional user classification data presented here. 
These variations could be found across all demographics, or they could be correlated with skill level, and additional work is needed to identify likely causes of these differences.
If tournament organizers endeavor may be tempted to artificially increase latency to balance across users, which would be similar to adding handicaps in traditional sports, and is not generally recommended.

\section{Conclusion}
Based on previous publications and the data we present here, we conclude that reducing the latency of a computer system is beneficial to FPS aiming performance, even at low latency levels. 
Certainly based on response time alone, we believe that the most competitive of players should try to have the lowest latency possible.
However, beyond this minimal advantage, latency does seem to offer benefits to users even in the sub-20 ms range, with our subjects demonstrating 182 ms of median task time improvement at an 8 ms change in latency.

While in many cases latency will not be the determining factor in which player "wins" an interaction, having competitors use computer systems with the same latency behavior removes this important variable from distorting the outcome of a match.
One option for reducing the impacts of varied latency on tournament outcomes would be establishing more rigorous latency standards for hardware, enforcing that all competitors are made aware of their system latency and how different hardware components impact that latency.
Alternatively identical systems could be provided by tournament hosts, guaranteeing a level playing field between competitors, at the expense of allowing competitors to choose their own (peripheral) hardware.
By better leveling the playing field of system latency in esports we hope to strengthen its competitive integrity.
We strongly encourage esports players, teams, and tournament organizers to monitor and control latency both in training and competition.

We recognize that players and teams may have reasons to make decisions that trade latency for some other benefit.
A player may greatly prefer a particular mouse grip for example, and a loss of latency may be seen as less important than a mouse supporting this grip.
A team may sign a sponsorship deal that includes hardware restrictions.
In these situations, we believe the decision makers should be aware of the aiming performance cost of increased latency and include it in their cost-benefit analysis.

%% file: main.bbl

\begin{thebibliography}{9}

%%% the Web address.

\ifx \showCODEN    \undefined \def \showCODEN     #1{\unskip}     \fi
\ifx \showDOI      \undefined \def \showDOI       #1{#1}\fi
\ifx \showISBNx    \undefined \def \showISBNx     #1{\unskip}     \fi
\ifx \showISBNxiii \undefined \def \showISBNxiii  #1{\unskip}     \fi
\ifx \showISSN     \undefined \def \showISSN      #1{\unskip}     \fi
\ifx \showLCCN     \undefined \def \showLCCN      #1{\unskip}     \fi
\ifx \shownote     \undefined \def \shownote      #1{#1}          \fi
\ifx \showarticletitle \undefined \def \showarticletitle #1{#1}   \fi
\ifx \showURL      \undefined \def \showURL       {\relax}        \fi
% invisible to TeX
\providecommand\bibfield[2]{#2}
\providecommand\bibinfo[2]{#2}
\providecommand\natexlab[1]{#1}
\providecommand\showeprint[2][]{arXiv:#2}

\bibitem[\protect\citeauthoryear{Cattan, Rochet-Capellan, Perrier, and
  B\'{e}rard}{Cattan et~al\mbox{.}}{2015}]%
        {cattan2015reducing}
\bibfield{author}{\bibinfo{person}{Elie Cattan}, \bibinfo{person}{Am\'{e}lie
  Rochet-Capellan}, \bibinfo{person}{Pascal Perrier}, {and}
  \bibinfo{person}{Fran\c{c}ois B\'{e}rard}.} \bibinfo{year}{2015}\natexlab{}.
\newblock \showarticletitle{Reducing Latency with a Continuous Prediction:
  Effects on Users' Performance in Direct-Touch Target Acquisitions}. In
  \bibinfo{booktitle}{\emph{Proceedings of the 2015 International Conference on
  Interactive Tabletops and Surfaces}} (Madeira, Portugal)
  \emph{(\bibinfo{series}{ITS '15})}. \bibinfo{publisher}{Association for
  Computing Machinery}, \bibinfo{address}{New York, NY, USA},
  \bibinfo{pages}{205–214}.
\newblock
\showISBNx{9781450338998}
\urldef\tempurl%
\url{https://doi.org/10.1145/2817721.2817736}
\showDOI{\tempurl}


\bibitem[\protect\citeauthoryear{Fitts and Peterson}{Fitts and
  Peterson}{1964}]%
        {fitts1964information}
\bibfield{author}{\bibinfo{person}{Paul~M Fitts} {and} \bibinfo{person}{James~R
  Peterson}.} \bibinfo{year}{1964}\natexlab{}.
\newblock \showarticletitle{Information capacity of discrete motor responses.}
\newblock \bibinfo{journal}{\emph{Journal of experimental psychology}}
  \bibinfo{volume}{67}, \bibinfo{number}{2} (\bibinfo{year}{1964}),
  \bibinfo{pages}{103}.
\newblock


\bibitem[\protect\citeauthoryear{Ivkovic, Stavness, Gutwin, and
  Sutcliffe}{Ivkovic et~al\mbox{.}}{2015}]%
        {ivkovic2015quantifying}
\bibfield{author}{\bibinfo{person}{Zenja Ivkovic}, \bibinfo{person}{Ian
  Stavness}, \bibinfo{person}{Carl Gutwin}, {and} \bibinfo{person}{Steven
  Sutcliffe}.} \bibinfo{year}{2015}\natexlab{}.
\newblock \bibinfo{booktitle}{\emph{Quantifying and Mitigating the Negative
  Effects of Local Latencies on Aiming in 3D Shooter Games}}.
\newblock \bibinfo{publisher}{Association for Computing Machinery},
  \bibinfo{address}{New York, NY, USA}, \bibinfo{pages}{135–144}.
\newblock
\showISBNx{9781450331456}
\urldef\tempurl%
\url{https://doi.org/10.1145/2702123.2702432}
\showURL{%
\tempurl}


\bibitem[\protect\citeauthoryear{Jota, Ng, Dietz, and Wigdor}{Jota
  et~al\mbox{.}}{2013}]%
        {jota2013howfast}
\bibfield{author}{\bibinfo{person}{Ricardo Jota}, \bibinfo{person}{Albert Ng},
  \bibinfo{person}{Paul Dietz}, {and} \bibinfo{person}{Daniel Wigdor}.}
  \bibinfo{year}{2013}\natexlab{}.
\newblock \bibinfo{booktitle}{\emph{How Fast is Fast Enough? A Study of the
  Effects of Latency in Direct-Touch Pointing Tasks}}.
\newblock \bibinfo{publisher}{Association for Computing Machinery},
  \bibinfo{address}{New York, NY, USA}, \bibinfo{pages}{2291–2300}.
\newblock
\showISBNx{9781450318990}
\urldef\tempurl%
\url{https://doi.org/10.1145/2470654.2481317}
\showURL{%
\tempurl}


\bibitem[\protect\citeauthoryear{Looser, Cockburn, and Savage}{Looser
  et~al\mbox{.}}{2005}]%
        {looser2005validity}
\bibfield{author}{\bibinfo{person}{Julian Looser}, \bibinfo{person}{Andy
  Cockburn}, {and} \bibinfo{person}{Joshua Savage}.}
  \bibinfo{year}{2005}\natexlab{}.
\newblock \showarticletitle{On the validity of using First-Person Shooters for
  Fitts' law studies}.
\newblock \bibinfo{journal}{\emph{People and Computers XIX}}
  \bibinfo{volume}{2} (\bibinfo{year}{2005}), \bibinfo{pages}{33--36}.
\newblock


\bibitem[\protect\citeauthoryear{MacKenzie and Ware}{MacKenzie and
  Ware}{1993}]%
        {mackenzie1993lag}
\bibfield{author}{\bibinfo{person}{I.~Scott MacKenzie} {and}
  \bibinfo{person}{Colin Ware}.} \bibinfo{year}{1993}\natexlab{}.
\newblock \showarticletitle{Lag as a Determinant of Human Performance in
  Interactive Systems}. In \bibinfo{booktitle}{\emph{Proceedings of the
  INTERACT '93 and CHI '93 Conference on Human Factors in Computing Systems}}
  (Amsterdam, The Netherlands) \emph{(\bibinfo{series}{CHI '93})}.
  \bibinfo{publisher}{Association for Computing Machinery},
  \bibinfo{address}{New York, NY, USA}, \bibinfo{pages}{488–493}.
\newblock
\showISBNx{0897915755}
\urldef\tempurl%
\url{https://doi.org/10.1145/169059.169431}
\showDOI{\tempurl}


\bibitem[\protect\citeauthoryear{Spjut, Boudaoud, Binaee, Kim, Majercik,
  McGuire, Luebke, and Kim}{Spjut et~al\mbox{.}}{2019a}]%
        {spjut2019latency}
\bibfield{author}{\bibinfo{person}{Josef Spjut}, \bibinfo{person}{Ben
  Boudaoud}, \bibinfo{person}{Kamran Binaee}, \bibinfo{person}{Jonghyun Kim},
  \bibinfo{person}{Alexander Majercik}, \bibinfo{person}{Morgan McGuire},
  \bibinfo{person}{David Luebke}, {and} \bibinfo{person}{Joohwan Kim}.}
  \bibinfo{year}{2019}\natexlab{a}.
\newblock \showarticletitle{Latency of 30 ms Benefits First Person Targeting
  Tasks More Than Refresh Rate Above 60 Hz}.
\newblock In \bibinfo{booktitle}{\emph{SIGGRAPH Asia 2019 Technical Briefs}}.
  \bibinfo{publisher}{ACM}, \bibinfo{address}{Brisbane, Australia},
  \bibinfo{pages}{110--113}.
\newblock


\bibitem[\protect\citeauthoryear{Spjut, Boudaoud, Binaee, Majercik, McGuire,
  and Kim}{Spjut et~al\mbox{.}}{2019b}]%
        {Spjut19FPSci}
\bibfield{author}{\bibinfo{person}{Josef Spjut}, \bibinfo{person}{Ben
  Boudaoud}, \bibinfo{person}{Kamran Binaee}, \bibinfo{person}{Alexander
  Majercik}, \bibinfo{person}{Morgan McGuire}, {and} \bibinfo{person}{Joohwan
  Kim}.} \bibinfo{year}{2019}\natexlab{b}.
\newblock \showarticletitle{FirstPersonScience: Quantifying Psychophysics for
  First Person Shooter Tasks}. In \bibinfo{booktitle}{\emph{UCI Esports
  Conference}}. \bibinfo{publisher}{UCI}, \bibinfo{address}{Irvine, CA}, 7.
\newblock


\bibitem[\protect\citeauthoryear{{Teather}, {Pavlovych}, {Stuerzlinger}, and
  {MacKenzie}}{{Teather} et~al\mbox{.}}{2009}]%
        {teather2009effects}
\bibfield{author}{\bibinfo{person}{R.~J. {Teather}}, \bibinfo{person}{A.
  {Pavlovych}}, \bibinfo{person}{W. {Stuerzlinger}}, {and}
  \bibinfo{person}{I.~S. {MacKenzie}}.} \bibinfo{year}{2009}\natexlab{}.
\newblock \showarticletitle{Effects of tracking technology, latency, and
  spatial jitter on object movement}. In \bibinfo{booktitle}{\emph{2009 IEEE
  Symposium on 3D User Interfaces}}. \bibinfo{publisher}{IEEE},
  \bibinfo{address}{Lafayette, LA, USA}, \bibinfo{pages}{43--50}.
\newblock
\urldef\tempurl%
\url{https://doi.org/10.1109/3DUI.2009.4811204}
\showDOI{\tempurl}


\end{thebibliography}
